\documentclass[twocolumn,prl,reprint]{revtex4-1}
\usepackage{amsfonts}
\usepackage{amsmath}
\usepackage{amssymb}
\usepackage{color}
\usepackage{graphicx}
\usepackage{bm}
\usepackage{comment}
\usepackage{esint}

\begin{document}

\title{Microwave spectroscopy of a weakly-pinned charge density wave
in a superinductor}

\author{Manuel Houzet}
\affiliation{Univ.~Grenoble Alpes, CEA, INAC-Pheliqs, F-38000 Grenoble, France}
\author{Leonid I. Glazman}
\affiliation{Department of Physics, Yale University, New Haven, CT 06520, USA}

\begin{abstract}
A chain of small Josephson junctions (aka superinductor) emerged recently as a high-inductance, low-loss element of superconducting quantum devices.  We notice that the intrinsic parameters of a typical superinductor in fact place it into the Bose glass universality class for which the propagation of waves in a sufficiently long chain is hindered by pinning. Its weakness provides for a broad crossover from the spectrum of well-resolved plasmon standing waves  at high frequencies to the low-frequency excitation spectrum of a pinned charge density wave. We relate the scattering amplitude of microwave photons reflected off a superinductor to the dynamics of a Bose glass. The dynamics at long and short scales compared to the Larkin pinning length determines the low- and high-frequency asymptotes of the reflection amplitude.
\end{abstract}

\maketitle

Interaction between particles gives rise to collective excitations in a many-body system. In the case of Coulomb interaction, these are the well-known plasma oscillation modes. The long-range interaction between particles confined to one or two dimensions (1D or 2D) may be cut-off by the polarizability of the surrounding medium. Translational invariance in 1D or 2D then results in the sound-like plasmon spectrum at low frequency. An ubiquitously present disorder breaks translational invariance and possibly affects the spectrum of low-frequency excitations. The competition between interaction and disorder sets the stage to a possible localization transition in a many-body system. This competition in 1D was addressed by means of perturbative renormalization group (RG) theory by Giamarchi and Schulz~\cite{Giamarchi1988}. Interaction in 1D can be characterized by a dimensionless parameter $K$ related to the magnitude of zero-point fluctuations of the particles' density (the classical limit is $K\to 0$). It turns out that the disorder potential is irrelevant at $K>3/2$ and the sound-like mode does exist down to the smallest wave vectors ($q\to 0$). However, at $K<3/2$ even an infinitesimally weak disorder results in localization, severely affecting the properties of 1D systems at long spatial scales. Attempts to understand the localized phase have led to the notion of Bose glass phase~\cite{Fisher1989} and to establishing its links to the pinned
vortices in superconductors~\cite{Larkin1970}, domain walls in magnets~\cite{Imry1975}, and ``classical'' charge-density waves in normal metals~\cite{Fukuyama1978}. 

In the localized phase, the static spatial order is destroyed on the scale exceeding the Larkin length, $R_\star$, which was first introduced~\cite{Larkin1970} for the collective pinning of vortices (these were modeled as classical particles). For weak pinning, $R_\star$ exceeds significantly the inter-particle distance. Elastic properties of the pinned system on length scales shorter than $R_\star$ are hardly affected by pinning. Respectively, in a non-dissipative system excitations with frequencies $\omega\gtrsim \omega_\star\equiv v/R_\star$  may still be approximated by propagating waves (here $v$ is the propagation speed)~\cite{Gorkov1977}. Pinning drastically changes the excitation spectrum at frequencies $\omega\ll\omega_\star$. The corresponding density of modes arises from the statistics of specific configurations of disorder supporting localized-in-space low-frequency excitations~\cite{Gorkov1977,Fukuyama1978,Feigelman1980,Feigelman1981,Aleiner1994,Fogler2002,Gurarie2002,Gurarie2003}. The found~\cite{Aleiner1994,Fogler2002,Gurarie2002,Gurarie2003} limiting low-frequency behavior of the density of modes (per unit volume) is $\nu(\omega)\propto\omega^4$.

Conductivity $\sigma(\omega)$, being sensitive to the frequency dependence of the transition matrix elements along with that for the density of modes, carries some information regarding the dynamics of charge density waves. It was predicted to have a maximum at $\omega\sim\omega_\star$, see, {\sl e.g.},~\cite{Fukuyama1978b,Chitra1998,Fertig1999,Fogler2000}. Experiments performed with a 2D electron gas in GaAs heterostructures qualitatively confirmed the predictions, but disagreed with them quantitatively, see \cite{Giamarchi2002} for a review.
In 1D, the physics of charge density waves was addressed in an experiment~\cite{Cedergren2017} where nonlinear current-voltage characteristics of Josephson-junction chains were studied. The tell-tale signature of the pinning was the appearance of dissipative current above a threshold voltage and a specific, systematic dependence of the threshold voltage on the parameters of a chain. The threshold voltage is related to the pinning energy at scale $R_\star$~\cite{Vogt2015,Vogt2016}, which is the basic property of the static pinning configuration.

\begin{figure}
\includegraphics[width=0.8\columnwidth]{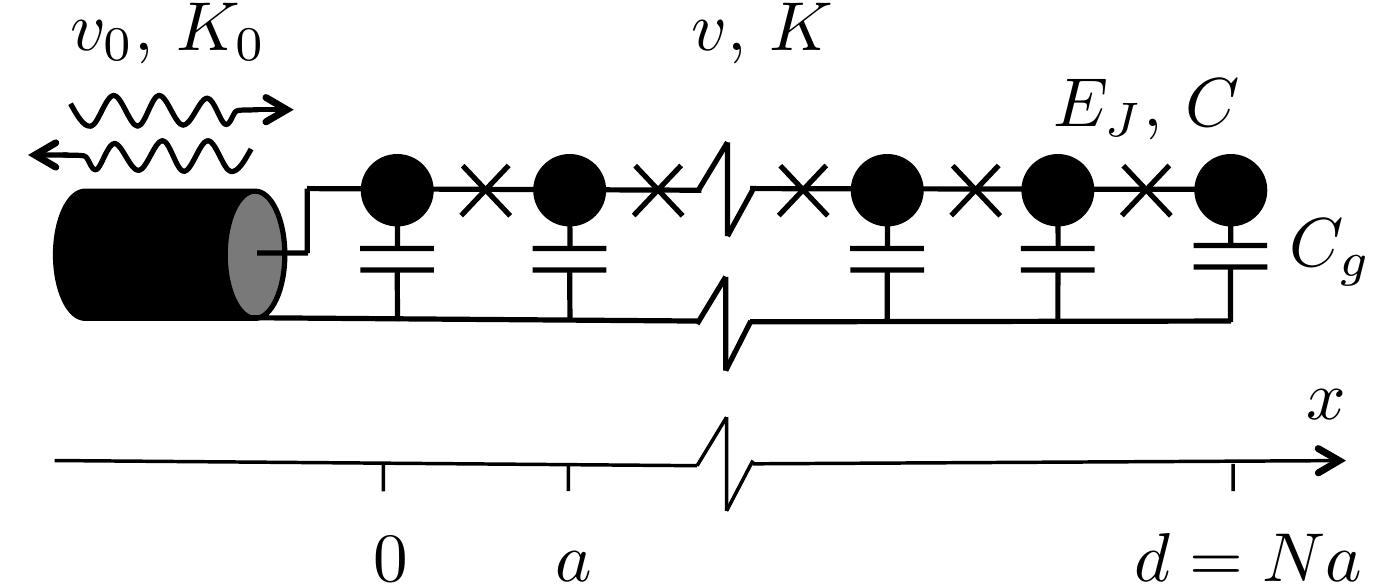}
\caption{\label{Fig:circuit} Microwave photons incident from a transmission line are reflected off a superinductor formed of $N$ Josephson junctions in series. }
\end{figure} 
In this work, we elucidate a way to study the dynamics of charge-density waves with a special type of Josephson-junction arrays, known as superinductors. Developed in the context of superconducting quantum devices~\cite{Manucharyan2009}, superinductors are linear elements combining high inductance with a small stray capacitance. An equivalent circuit of a superinductor is shown in Fig.~\ref{Fig:circuit}. Large inductance and linearity are achieved by making the number of  junctions large, $N\gg 1$, and quantum fluctuations of the phase across a single junction small, $E_J/E_C\gg 1$, while having small stray capacitance calls for a very small ratio $E_C/E_g$ [here $E_J$ is the Josephson energy of a single junction, $E_g=4e^2/(2C_g)$ and $E_C=4e^2/(2C)$ are the charging energies for an extra Cooper pair associated, respectively, with the superconducting island's stray capacitance $C_g$ and the junction capacitance $C$; the parameters were $N\sim 10^2$, $E_J/E_C\sim 20$, and $E_C/E_g\sim 10^{-4}$ in the experiment~\cite{Manucharyan2009}]. The product of $E_J/E_C$ and $E_C/E_g$ turns out to be small, resulting in $K<1$, so nominally superinductors are insulators. However, the amplitude of quantum phase slips and therefore the pinning potential are exponentially small, $\exp(-\sqrt{32 E_J/E_C})\ll 1$. Unless $N\gg 1$ compensates for that smallness, phase slips are rare and the superinductor faithfully performs its inductance function in a circuit~\cite{Manucharyan2009,Masluk2012,Bell2012}. Recently, even longer chains with $N\sim 10^4$ were developed~\cite{Kuzmin2018}  for which the statistics of quantum phase fluctuations allows a finite density of quantum phase slips to appear. This, in turn, enables weak pinning of charge density waves. Realization of the pinning potential depends on the random background charges in the environment of the chain. These are slowly fluctuating in time~\cite{Riste2013}, providing a tool for the ensemble averaging of observables. We evaluate the most accessible one, which is the ensemble-averaged reflection amplitude off a chain, $\langle r(\omega)\rangle$, find its relation to the local density of states of excitations, and predict the low- and high-frequency asymptotes of $\langle r(\omega)\rangle$.
\\

We model the superinductor with the Hamiltonian
\begin{equation}
\label{eq:H}
{H}_\text{chain} =\frac12\sum_{nm}Q_nC^{-1}_{nm}Q_m
-E_J
\sum_n\cos(\varphi_n-\varphi_{n+1})\,,
\end{equation}
where $\varphi_n$ and $Q_n$ are canonically conjugated phase and charge of each superconducting island along the chain, $[\varphi_n,Q_m]=2ei\delta_{n,m}$. % with electric charge $e$.
The first term in Eq.~\eqref{eq:H} describes the electrostatic 
coupling with elements $C_{nn}=(2C+C_g)$ and $C_{nn\pm 1}=C$ of the capacitance matrix,
the second term describes the Josephson coupling between successive islands. Hamiltonian \eqref{eq:H} can be used if the temperature, charging, and Josephson energy are smaller than the superconducting gap. The typically small stray capacitance $C_g\ll C$ corresponds to a large charge screening length, $\ell_{\mathrm{sc}}=a\sqrt{C/C_g}\gg a $, where $a$ is the unit cell length, see Fig.~\ref{Fig:circuit}.

In harmonic approximation, Hamiltonian~\eqref{eq:H} yields the dispersion relation $\omega(q)=v|q|/\sqrt{1+(vq/\Omega)^2}$, where $v = a \sqrt{2E_J E_g}/\hbar$ is the plasmon speed with wave vectors $q$ below $\ell_{\rm sc}^{-1}$ or, equivalently, with $\omega(q)$ below the single-junction plasma frequency $\Omega=\sqrt{2E_JE_c}/\hbar$. Modes with $\omega(q)\approx v|q|$ are adequately described by a harmonic string Hamiltonian,
\begin{eqnarray}
H_0=\int dx \left[ \frac{\pi v K}{2\hbar}\Pi^2 +\frac{\hbar v}{2\pi K}(\partial_ x\theta)^2\right]
\label{eq:H-quad}
\end{eqnarray}
acting on states with energies within the bandwidth $\sim\hbar\Omega$. Here $\theta$ and $\Pi=-(\hbar /\pi)\partial_x\varphi$ are two canonically conjugated fields, $[\theta(x), \Pi(x')]=i\hbar \delta(x-x')$, where  $\varphi$ and $\rho=-(1/\pi)\partial_x\theta$ are the coarse-grained phase and Cooper-pair density along the chain, respectively. The parameter $K=\pi\sqrt{E_J /(2E_g)}$ is related to the low-frequency impedance $Z$ of the chain, $K=\pi\hbar/(4e^2Z)$.

Quantum phase slips allow for jumps of the phase differences $\varphi_n-\varphi_{n+1}$ between the minima of the Josephson energy in Eq.~\eqref{eq:H}. Rare phase slips can be accounted for by adding \cite{Gurarie2004} a perturbative term to the Hamiltonian~\eqref{eq:H-quad},
\begin{equation}
\label{eq:Hdis}
H =H_0- \frac\lambda a\int dx \cos(2\theta+\chi)\,.
\end{equation}
Operators $e^{\pm 2i \theta(x)}$ appearing in Eq.~\eqref{eq:Hdis} create $\pm 2\pi$-kinks at position $x$ in the field $\varphi(x)$. The classical  field $\chi(x)$ leads to the  Aharonov-Casher effect in the probability amplitudes of phase slips~\cite{Ivanov2001,Friedman2002}. 

Field $\chi(x)=2\pi  \int^xdx' \rho_\mathrm{b}(x ')$ is random, $\rho_\mathrm{b}(x)$ is the density of coarse-grained offset charges.
A maximal disorder corresponds to offset charges fluctuating independently and randomly in each superconducting island. It yields a Gaussian correlator
\begin{equation}
\langle \cos\chi(x) \cos\chi (x')\rangle=\frac 12 a\delta(x-x') \,,
\label{eq:correl-cont2}
\end{equation}
where $\langle\cdots\rangle$ stands for disorder averaging, for the field $\cos\chi$ (a similar relation holds for the field $\sin\chi$) on spatial scales larger than $\ell_\mathrm{sc}$.
Furthermore, a recent experiment~\cite{Riste2013} reported timescale $t_c\sim 1{\rm min}$ for the offset charge fluctuations of a single island. Extrapolating their results to a chain yields a timescale $t_c/N$ for scrambling the Aharonov-Casher phase in a chain of $N$ junctions.

Due to the separation of scales in the plasmon spectrum at $C_g/C\ll 1$ and the known full solution~\cite{Korshunov1989,Matveev2002} of the phase-slip problem at $C_g/C=0$, it is possible, in contrast to the phenomenological treatments~\cite{Gurarie2004,Bard2017}, to derive Hamiltonian \eqref{eq:Hdis} and evaluate $\lambda$ in terms of microscopic parameters,
\begin{equation}
\label{eq:lambda}
\lambda=\frac{8}{\sqrt{\pi}} (2E_J^3E_c)^{1/4}e^{-\sqrt{32E_J/E_c}} \quad\text{at}\quad E_J\gg E_c\,.
\end{equation}
Once $\lambda$ is known, the ${\cal O}(1)$ uncertainty in the bandwidth $\sim\hbar\Omega$ translates, at $K\ll 1$, into a negligible ${\cal O}(K)$ uncertainty in the low-frequency ($\omega\ll\Omega$) observables which we aim to evaluate.

In the thermodynamic limit (infinitely long chain), the  perturbative RG flow associated with the Hamiltonian \eqref{eq:Hdis} is given by the Giamarchi-Schulz scaling~\cite{Giamarchi1988}
\begin{equation}
\label{eq:RGdis1}
\frac{dD(\Lambda)}{dl}
\simeq(3 -2 K)D(\Lambda)\,.
\end{equation}
The unitless function $D(\Lambda)$ here describes the evolution of the phase slip probability in the process of coarse-graining, $\Lambda$ is the running momentum cutoff, and $dl =- d\Lambda/\Lambda$. The initial condition for Eq.~(\ref{eq:RGdis1}) is $D(\Lambda_0)=v\lambda^2/(a\Omega^3)$ with $\lambda$ of Eq.~\eqref{eq:lambda} and $\Lambda_0\sim \Omega/v$.
The first term in the prefactor $3-2K$ in Eq.~\eqref{eq:RGdis1} corresponds to the rescaling of two time but only one space coordinates (as the disorder is local in space and approximately static on the plasmon flight time, $d/v\ll t_c$), while each term $K$ accounts for the renormalization of the phase slip amplitude $\lambda$ at a single Josephson junction in the presence of an ohmic bath of long-wavelength plasmons. Equation~\eqref{eq:RGdis1} signals a transition between a superfluid phase ($K>K_c$) and a Bose glass ($K<K_c$) at $K_c = 3/2$. At small fugacity $\exp(-\sqrt{32E_J/E_c})$, see Eq.~\eqref{eq:lambda}, one may disregard the renormalization~\cite{Giamarchi1988} of $K$ in  finding the transition condition, $E_g =(2\pi^2/9)E_J$.

In the classical limit, corresponding to $K\to 0$ and $\hbar\to 0$, while the ratio $K/\hbar$ is maintained fixed, one may neglect the kinetic term in Eq.~\eqref{eq:Hdis} and estimate the Larkin length for the static pinning of the charge density from energy arguments: On one hand, a deformation of a static field $\bar \theta(x)$ by $2\pi$ over the length $R$ costs an elastic energy $E_{\rm el}\sim \hbar v/(KR)$. On the other hand, for a constant field $\bar\theta$, the disorder-averaged pinning energy vanishes, while its typical value for a given disorder configuration is estimated as $E_{\rm p}\sim \lambda\sqrt{R/a}$ using the correlator~\eqref{eq:correl-cont2}. The pinning energy dominates the elastic one if $R>R_\star$, where $R_\star=a\{\hbar v/[aK\lambda(R_\star)]\}^{2/3}$. Using here the renormalized phase slip amplitude $\lambda(R)=\lambda(R/\ell_\mathrm{sc})^{-K}$ instead of its bare value~\eqref{eq:lambda} allows us to account for quantum fluctuations on short length scales $R\lesssim R_\star$. Solving then for $R_\star$, we find the generalized Larkin length, 
\begin{equation}
\label{eq:LarkLength-Kfinite}
R_\star=a\left(\frac{\hbar \Omega}{K\lambda}\right)^{2/(3-2K)}\left(\frac C{C_g}\right)^{(1-K)/(3-2K)}\,.
\end{equation}
Remarkably, $1/R_\star$ coincides with the momentum scale at which the perturbative RG breaks down, $D(1/R_\star)\sim 1$.

We are now ready to formulate the problem of the elastic scattering of microwave photons off a Josephson-junction chain of a finite length $d$. For this, we consider the same Hamiltonian \eqref{eq:Hdis} but with spatially nonuniform parameters such that it describes a transmission line with plasmon speed $v_0$ and impedance $Z_0=\pi\hbar/(4e^2K_0)$ (and without phase slips, $\lambda=0$) at $x<0$, and the superinductor at $0<x<d$, see Fig.~\ref{Fig:circuit}. The equations of motion derived from Eq.~\eqref{eq:Hdis} yield $\ddot \theta=-v^2\partial_x^2\theta+(2\lambda/a)\sin(2\theta+\chi)$ together with the boundary conditions expressing the continuity of current, $\partial_t\theta(0^+,t)=\partial_t\theta(0^-,t)$, and voltage, 
$({v}/{K})\partial_x\theta(0^+,t)=({v_0}/{K_0})\partial_x\theta(0^-,t)$, at the interface between the waveguide and the chain, as well as the absence of current, $\partial_t\theta(d,t)=0$ at the other end of the chain. Taking the classical limit, we can now define the reflection amplitude $r(\omega)$ at frequency $\omega$, such that the solution of these equations is expressed as $\theta(x,t)=\bar\theta(x)+\psi(x)e^{-i\omega t}$, where $\bar\theta(x)$ is a static charge density that minimizes the (classical) energy, and $\psi(x)$ describes small oscillations around it. The linearized equation of motion takes a form similar to the Schr\"odinger equation,
\begin{equation}
\label{eq:schrod}
\omega^2 \psi=-v^2\partial_x^2\psi+V(x)\psi
\quad\mathrm{at}\quad 0<x<d\,,
\end{equation} 
with the normalization condition
\begin{equation}
\psi(x)=e^{i \omega x/v_0}+r(\omega)e^{-i \omega x/v_0}\qquad\mathrm{at}\qquad x<0\,,
\end{equation}
and potential $V(x)=(4\pi { K\lambda v}/{\hbar a})\cos(2\bar\theta(x)+\chi(x))$ which is determined both by the offset charge disorder and the static charge density $\bar \theta(x)$.

The impedance of typical waveguides is of the order of the vacuum impedance, $Z_\mathrm{vac}\approx 377\,\Omega$; thus, $K\ll K_0.$ Using this, we can find an expression for $r(\omega)$ in terms of the properties of the chain valid at arbitrary disorder configuration. Namely, we observe that at $K/K_0\to 0$, the chain is disconnected from the waveguide. Thus, it admits a set of discrete bound states with eigenfrequencies $\omega_n$ and eigenfunctions $\psi_n(x)$, which satisfy the Schr\"odinger equation~\eqref{eq:schrod} with boundary conditions $\partial_x\psi(0_n^+)=0$ and $\psi_n(d)=0$. We also impose the normalization condition $\int_0^ddx\,\psi^2_n(x)=1$. At small but finite $K/K_0$, these solutions become quasi-bound states: they emit plasmons in the waveguide, such that $\psi(x<0)=\psi(0^-)e^{-i\omega_n x/v_0}$ with $\psi(0^-)=\psi_n(0^+)$ according to the boundary condition. The energy stored in the bound state is $E_n= \omega_n^2/(\pi vK)$, while the energy emitted in the waveguide is characterized by the Poynting vector $P= v_0\omega_n^2/(\pi v_0K_0)\psi^2_n(0^+)$. The rate of energy loss, $\Gamma_n\equiv P/E_n=(K/K_0)v \psi_n^2(0^+)$, gives the level width of the quasi-bound state. We can now use the Breit-Wigner formula to account for the contribution of all quasi-bound states, $r(\omega)=-1+i\sum_n\Gamma_n/(\omega-\omega_n+i\Gamma_n/2)$, assuming that {$\Gamma_n\ll |\omega_n-\omega_{n+ 1}|$}, where $|\omega_n-\omega_{n+1}|$ is the level spacing between successive bound states. 
Let us now introduce a Green function of the chain, which solves 
\begin{equation}
\label{eq:green}
\left[\omega^2+v^2\partial_x^2 -V(x)\right]G(x,x';\omega)=\pi v  \delta(x-x')
\end{equation}
with boundary conditions
$\partial_x G(0,x';\omega)=G(x,d;\omega)=0$. Introducing the local plasmon density of states 
\begin{equation}
\label{eq:local-dos2}
\nu(x,\omega)=-\frac {2\omega}{\pi^2 v}  \text{Im}\,G(x,x;\omega)\,,
\end{equation}
defining $\nu_0=1/(\pi v)$, and using the complete basis of normalized eigenstates $\psi_n(x)$ to express the Green function allows us to find a relation between the real part of the reflection amplitude and the density of states at the edge of the chain,
\begin{equation}
\label{eq:local-dos}
r'(\omega)=-1+\frac{K}{K_0} \frac{\nu(x=0,\omega)}{\nu_0}\,.
\end{equation} 

The random charge realization changes on time scale $t_c/N$ far exceeding the typical plasmon propagation time $d/v$. 
For a static disorder, $r(\omega)$ is the sum of narrow peaks corresponding to plasmon resonances in the finite-size chain. We now evaluate its disorder average, assuming that the measurement time exceeds $t_c/N$, thus facilitating the averaging. 

At large frequencies, $\omega\gg\omega_\star$, the plasmon wavelength $v/\omega$ is much smaller than $R_\star$. This justifies neglecting the spatial variations of the static field $\bar\theta$ appearing in the disorder potential in Eq.~\eqref{eq:schrod}, which then becomes Gaussian. The potential induces both forward- and back-scattering characterized by a frequency-dependent mean-free path evaluated, within Born approximation, as 
\begin{equation}
\label{eq:mfp}
\ell(\omega)=v \tau(\omega)=R_\star(\omega/\omega_\star)^2\,.
\end{equation}
Ignoring the back-scattering for a while, we readily find that the local density of states is expressed as $\nu(0,\omega)=1/d\sum_n\delta(\omega-\omega_n)$, where $\omega_n=(n+1/2)\Delta+\delta_n$ corresponds to a spectrum of plasmon resonances nominally spaced by $\Delta=\pi v/d$ and randomly shifted by $\delta_n\approx 1/(2d\omega_n)\int_0^ddxV_0(x)$, where $V_0$ is the ``smooth'' component of $V$. Gaussian average over $V_0$ yields an ``inhomogeneous broadening'' of the peaks in $\langle\nu(0,\omega)\rangle$, which acquire a Gaussian lineshape.

Accounting for both forward- and back-scattering in the evaluation of  $\langle\nu(0,\omega)\rangle$ at large frequency $\omega\gg\omega_\star$ can be performed with the Fokker-Planck method,  see Sec.~I of the Supplemental Material (SM)~\cite{SM};
it only modifies the width of the Gaussian lineshapes, compared with the forward-scattering case. The result is
\begin{equation}
\label{eq:resutat-2}
\langle \nu(0,\omega)\rangle=2\nu_0\sum_m\frac{\Delta}{
\sqrt{2\pi }\delta(\omega)}e^{-\left[\omega -(m+1/2)\Delta\right]^2/[2\delta^2(\omega)]}\,,
\end{equation}
with the frequency-dependent width of the resonances,
\begin{equation}
\delta(\omega)=\Delta\cdot\frac{\omega_{\rm cr}}{\omega}
\quad\mathrm{and}\quad
\omega_{\rm cr}={\sqrt{3}}\omega_\star\left(\frac{d}{R_\star}\right)^{1/2}\,.
\end{equation}
The Gaussian-shaped resonances are well-separated as long as $\delta(\omega)\ll\Delta$, corresponding to the frequency range $\omega\gg \omega_{\rm cr}$ or, equivalently, for a chain's length $d\ll\ell(\omega)$.

At frequencies below the crossover, $\omega\lesssim \omega_{\rm cr}$, the resonances overlap, gradually suppressing the amplitude of oscillations of $r(\omega)$. Using the Poisson summation formula to transform Eq.~\eqref{eq:resutat-2}, we find 
\begin{equation}
\label{eq:resultat-poisson}
\langle \nu(0,\omega)\rangle=2\nu_0\left[1-2\cos\left({2\pi \omega}/{\Delta}\right)e^{-2(\pi \omega_{\rm cr}/{\omega})^2}\right]
\,
\end{equation}
in the frequency range $\omega_\star\ll\omega\ll \omega_\text{cr}$.
The frequency-independent part of Eq.~\eqref{eq:resultat-poisson} is the end density of states of a harmonic, half-infinite chain ($\lambda\to 0$, $d\to\infty$). 
The other terms account for small oscillations with the frequency; their period is $\Delta$, and their amplitude is exponentially suppressed, $\exp\{-2(\pi \omega_{\rm cr}/{\omega})^2\}=\exp\{-6\pi^2d/\ell(\omega)\}$. The amplitude decreases and reaches values $\sim\exp\{-{\rm const}\cdot (d/R_\star)\}$ with the frequency decreasing towards $\omega_\star$.
The oscillatory terms are exponentially small because the phase of the plasmon becomes scrambled on the length scale $\ell(\omega)$ much shorter than the length $d$ of the chain. 

The leading term in Eq.~(\ref{eq:resultat-poisson}) is independent of frequency, as at $\omega\gg\omega_\star$ the adjustment of the static configuration $\bar\theta(x)$ to the external charge disorder realization is not important.
In contrast, at $\omega\lesssim\omega_\star$ the plasmon wavelength becomes of the order of the correlation length of $\bar\theta(x)$. In this limit, the oscillatory part of $\langle r(\omega)\rangle$ remains exponentially small, but -- in addition -- the leading term in Eq.~(\ref{eq:resultat-poisson}) becomes a function of $\omega/\omega_\star$. The $\omega/\omega_\star\to 0$ asymptote of that function is a power-law with a universal exponent,
\begin{equation}
\label{eq:Dos-w4}
\langle\nu(x=0,\omega)\rangle = C\nu_0 (\omega/\omega_\star)^4\quad\text{at}\quad\omega/\omega_\star\ll 1\,.
\end{equation}
To see the universality of the exponent, we consider a sequence of models bridging the limits of weak and strong pinning, and then apply the ideas~\cite{Gorkov1977,Aleiner1994,Fogler2002,Gurarie2002,Gurarie2003} developed for the density of states in the {\it bulk} to derive the {\it edge} property $\langle\nu(x=0,\omega)\rangle$. Then, we find the constant $C\approx 0.032$ in Eq.~(\ref{eq:Dos-w4}) by a numerical simulation.    

To motivate Eq.~(\ref{eq:Dos-w4}), we may assume the chain to be half-infinite, as the boundary condition at $x=d$ should not matter due to the wave localization. The low-frequency plasmon spectrum will be contributed by the soft oscillation modes in special, barely stable configurations of disorder built in the vicinity of the $x=0$ edge of the chain. Next, we generalize the considered-so-far Gaussian model of disorder in continuum by substituting Hamiltonian \eqref{eq:Hdis} and correlator \eqref{eq:correl-cont2} with
\begin{equation}
H =H_0- \frac\lambda {\sqrt{ac}}\sum_j\cos(2\theta(x_j)+\chi_j)\,,
\end{equation}
where $\chi_j$ and $x_j$ are, respectively, the random phases associated with and random locations of discrete impurities. By varying their density $c$  from a large value down to $c\ll 1/a$ one crosses over between the limits of weak collective and strong individual-impurities pinning. The latter limit is amenable to the analytical treatment~\cite{Gorkov1977,Aleiner1994}.

Infinitely-strong pinning on separate impurities reduces the system to a sequence of independent segments~\cite{Gorkov1977}. Finite-strength impurities allow for special configurations carrying soft excitations with arbitrarily-low frequency, as was noticed in~\cite{Aleiner1994}. Closely following that work, we consider the needed three-impurity configurations in the vicinity of $x=0$. The only difference from~\cite{Aleiner1994} is that we impose the boundary condition $\partial_x\bar\theta(0)=0$ on the static charge distribution. Following~\cite{Aleiner1994}, we find impurity configurations resulting in the low-frequency local modes of the pinned elastic string. It is tedious but straightforward (see Sec.~II of SM~\cite{SM}) to show that the boundary condition at $x=0$ does not affect the $\omega\to 0$ asymptote of the density of states, $\langle\nu(\omega)\rangle\propto\omega^4$. The proportionality coefficient here depends on $c$, but we do not expect a phase transition to a different functional form upon reducing $c$.
As argued in~\cite{Fogler2002,Gurarie2002,Gurarie2003}, the functional form of the asymptote is universal and remains the same in the limit of weak pinning, which is of direct interest in the context of this work.

\begin{figure}
\includegraphics[width=0.9\columnwidth]{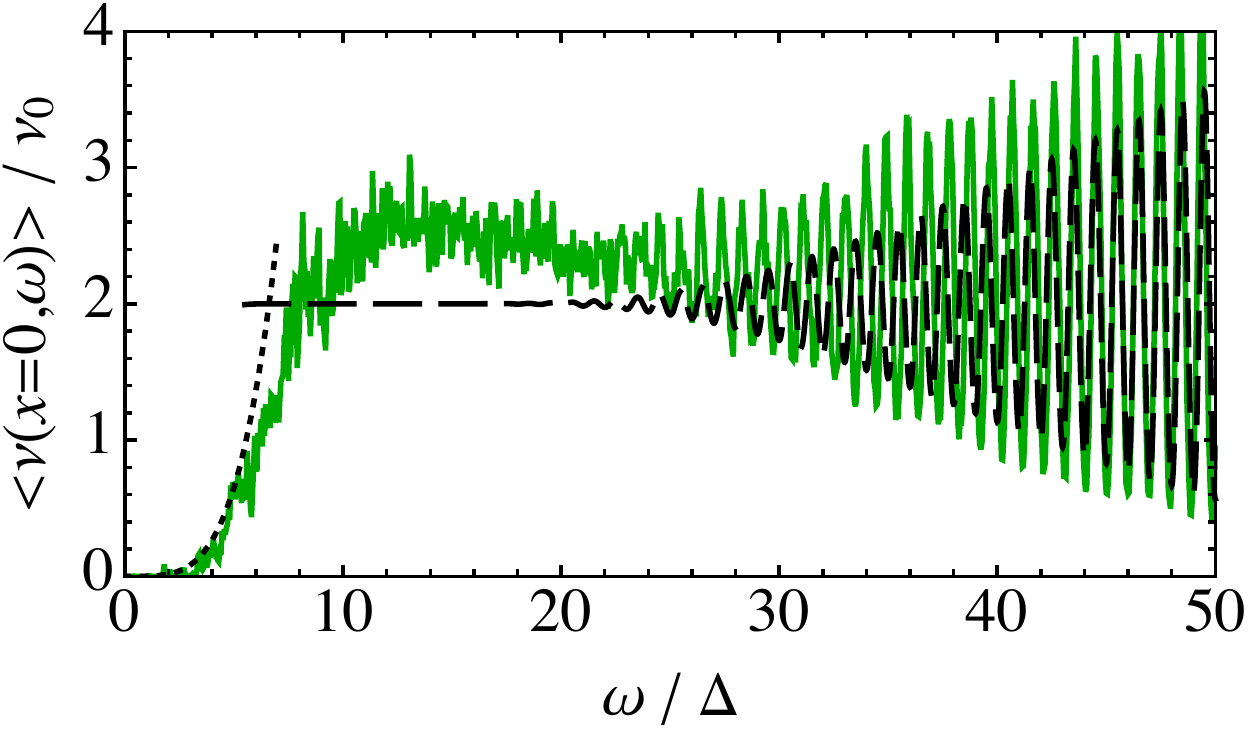}
\caption{\label{Fig:numerics} Frequency dependence of the edge density of states, numerically evaluated and averaged over 5000 disorder configurations in a chain of length $d=7.4 R_\star$ (straight line). For comparison, analytically obtained asymptotes, Eq.~\eqref{eq:resutat-2} (dashed line) and Eq.~\eqref{eq:Dos-w4} (dotted line), are also plotted.}
\end{figure}

The considerations that led to Eq.~\eqref{eq:Dos-w4} are substantiated -- and the proportionality coefficient in it is found -- in a numerical simulation presented in Sec.~III of SM~\cite{SM}, and whose result is illustrated in Fig.~\ref{Fig:numerics} covering a broad range of frequencies.

Using Eq.~\eqref{eq:local-dos} and the results \eqref{eq:resutat-2}, \eqref{eq:resultat-poisson}, and \eqref{eq:Dos-w4}, we can now predict the overall evolution of the photon reflection amplitude with the increase of frequency. The microwave photons are fully reflected with $\langle r'\rangle\approx -1$ at $\omega\ll\omega_\star$; according to Eq.~(\ref{eq:Dos-w4}), the average reflection increases approaching $\langle r'\rangle\approx -1+2K/K_0$ at $\omega\sim\omega_\star$. Upon further increase of $\omega$, the prominence of the spatial resonances in $\langle r'(\omega)\rangle$ raises, and they become well-resolved at $\omega\gg\omega_\text{cr}$, see Eqs.~(\ref{eq:resutat-2}) and (\ref{eq:resultat-poisson}). In the latter regime, the width of resonances translates into the quality factor $Q=\omega/[(4\ln 2) \delta(\omega)]\propto \omega^2$. We emphasize that the $Q$-factor of the resolved resonances, according to our theory, comes from the ensemble averaging applied to the elastic plasmon propagation. We note in passing, that the inelastic scattering, considered in~\cite{Bard2018,Wu2018} in the context of plasmon decay at $d\to\infty$, results in a minor contribution to $1/Q$ of well-resolved spatial resonances at finite $d$, see Sec.~IV of SM~\cite{SM}. 

In conclusion, microwave photon scattering off a superinductor may open a new way to study pinning in a one-dimensional quantum system. This work was devoted to the theory of reflection amplitude in the limit of small quantum fluctuations ($K\ll 1$). The reflection amplitude was recently measured~\cite{Kuzmin2018} for a variety of superinductors; in a qualitative agreement with our predictions, the increase of the $Q$-factor with the microwave frequency was seen for samples with the highest probability of quantum phase slips. 

%\end{document}

\begin{acknowledgments}
We acknowledge stimulating discussions with V. Manucharyan. This work is supported by the DOE contract DEFG02-08ER46482 (LG), and by the ARO grant W911NF-18-1-0212, a Google gift to Yale University, the European Union's FP7 programme through the Marie-Sk\l odowska-Curie Grant Agreement 600382, and ANR through grant No.~ANR-16-CE30-0019 (MH). 
\end{acknowledgments}


\begin{thebibliography}{99}
\bibitem{SM}
Details on the derivation of Eq.~\eqref{eq:resutat-2} including back-scattering, the prefactor of Eq.~\eqref{eq:Dos-w4} in the strong-pinning regime, the numerics, and the smallness of inelastic scattering rate can be found in the SM.
\end{thebibliography}
\end{document}

\title{Supplemental Material for:
\\``Microwave spectroscopy of a weakly-pinned charge density wave
in a superinductor''}

%\author{Manuel Houzet}
%\affiliation{Univ.~Grenoble Alpes, CEA, INAC-Pheliqs, F-38000 Grenoble, France}
%\author{Leonid I. Glazman}
%\affiliation{Department of Physics, Yale University, New Haven, CT 06520, USA}
%\begin{abstract}
%\end{abstract}
\maketitle

\section{Semiclassics}

Here we include back-scattering processes in the derivation of Eq.~(16) in the main text for the local density of states at large frequency, $\omega\gg\omega_\star$; we find that it only modifies the expression~(17) for $\omega_\text{cr}$ (the factor $\sqrt{3}$ would be replaced with $\sqrt{2}$ if only forward-scattering processes were present). Namely, at $\omega\gg\omega_\star$, we use the analogy with the one-dimensional Schr\"odinger equation for a particle in a Gaussian disorder potential to evaluate the averaged local plasmon density of states that appears in Eq.~(12) in the main text. For this, we employ the Fokker-Planck method reviewed in \cite{Lifshitz1988}.

We may express the Green function that solves Eq.~(10) in the main text, together with boundary conditions, as
\begin{equation}
G(x,x';\omega)=\frac {\pi } {vW} \left[\Psi_-(x)\Psi_+(x')\Theta(x'-x)+\Psi_-(x')\Psi_+(x)\Theta(x-x')\right]\,.
\label{eq:GFclassical}
\end{equation}
Here $\Psi_+$ and $\Psi_-$ are two solutions of the differential equation
\begin{equation}
\label{eq:diff}
\omega^2 \Psi_\pm(x)=-v^2\partial_x^2 \Psi_\pm(x)+V(x)\Psi_\pm(x)
\end{equation}
with boundary conditions $\partial_x \Psi_-(0)=0$ and $\Psi_-(0)=1$, and $\partial_x \Psi_+(d)=1$ and $\Psi_+(d)=0$, respectively,
$W$ is a spatially-independent Wronskian,
\begin{equation}
W=\Psi_-(x)\partial_x\Psi_+(x)-\partial_x\Psi_-(x)\Psi_+(x)\,,
\end{equation}
and $\Theta(x)$ is the Heaviside function [with $\Theta(0)=1/2$]. 
Using the Wronskian and the boundary conditions, we find from Eq.~\eqref{eq:GFclassical} 
\begin{equation}
\label{eq:G00}
G(0,0;\omega)=\frac{\pi }v\frac{\Psi_+(0)}{\partial_x\Psi_+(0)}\,.
\end{equation}
We further define four functions $\rho_\pm$ and $\phi_\pm$ such that 
%\begin{subequations}
\begin{equation}
\label{eq:parametrize}
\Psi_\pm(x)=\rho_\pm(x)\sin\phi_\pm(x) \quad\text{and}\quad 
\partial_x\Psi_\pm(x)=(\omega/v) \rho_\pm(x)\cos\phi_\pm(x)\,.
\end{equation}
%\end{subequations}
Equation \eqref{eq:diff} and its boundary conditions then read equivalently
\begin{equation}
\label{eq:neq-eq}
\partial_x \rho_\pm=-\frac{V}{2v\omega}\sin2\phi_\pm \quad\mathrm{and}\quad
\partial_x \phi_\pm=\frac\omega v-\frac{V}{v\omega}\sin^2\phi_\pm
\end{equation}
with boundary conditions $\rho_+(d)=v/\omega$ and $\phi_+(d)=0$, and $\rho_-(0)=1$ and $\phi_-(0)=\pi/2$, respectively.
Inserting Eq.~\eqref{eq:parametrize} into Eq.~\eqref{eq:G00}, we relate 
the statistics of 
\begin{equation}
\label{eq:G00-param}
G(0,0;\omega)=\frac{\pi }\omega\tan\phi_+(0)
\end{equation}
with that of the solution $\phi_+$ of the relevant pair of Eqs.~\eqref{eq:neq-eq}.

In general, $G(0,0;\omega)$ is complex. Its real part readily follows from Eq.~\eqref{eq:G00-param}. In order to derive its imaginary part, one should include a small imaginary part to the frequency (in the upper complex plane), $\omega\to \omega+i0^+$, which in turn adds a small imaginary part with the same sign to $\phi_+$. Using Eq.~(13) in the main text, we can then relate the statistical properties of the local density of states to those of $\phi_+(0)$,
\begin{equation}
\label{eq:res-rvdG3}
\nu(x=0,\omega)
 =-\frac 2 v \sum_m\delta\left(\phi_+(0)-(m+\frac12)\pi\right)\,.
\end{equation}

To proceed further, we use the assumption of large frequency to treat the term $\propto V$ in Eq.~(\ref{eq:neq-eq}) as a perturbation,
and decompose $\phi_+(x)=\omega (x-d)/v+\zeta(x)$ , where $\zeta(x)$ is a slowly varying function on the scale of $v/\omega$. 
Furthermore, the disorder potential contains three relevant terms for the spatial variation of $\zeta(x)$,
\begin{equation}
\label{eq:semiclassical-disorder}
V(x)=V_0(x)+V_1(x)e^{2i\omega (x-d)/v}+V^*_1(x)e^{-2i\omega(x-d)/v}\,,
\end{equation}
where the slowly varying (on the scale of $v/\omega$) components $V_0(x)$ and $V_1(x)$ describe forward- and back-scattering, and have Gaussian averages,
\begin{equation}
\label{eq:correl-reduced-2}
\langle V_0(x)V_0(x')\rangle=\langle V_1(x)V^*_1(x')\rangle= 8 \pi^2R_\star\omega_\star^4\delta(x-x')\,.
\end{equation}
Equation~\eqref{eq:correl-reduced-2} reproduces the correlator for $V= (4\pi K\lambda v/\hbar a)\cos(2\bar\theta+\chi)$, in which the spatial variations of $\bar\theta$ are safely ignored at $\omega\gg\omega_\star$ (see discussion in main text) and the correlator for $\cos\chi$ is given by Eq.~(4) in the main text. Then retaining only the slowly varying components in Eq.~(\ref{eq:neq-eq}), we find
\begin{equation}
\label{eq:neq-eq3}
\partial_x \zeta=-\frac 1{2v\omega}V_0+ \frac 1{4v\omega}\left[V_1^*e^{2i\zeta}+V_1e^{-2i\zeta}\right]
\end{equation}
with the boundary condition $\zeta(d)=0$. Equation~\eqref{eq:neq-eq3} is a Wiener process whose Fokker-Planck equation reads~\cite{Lifshitz1988}
\begin{equation}
\label{eq:const-diff}
\frac{\partial P}{\partial x}={\cal D}\frac{\partial ^2 P}{\partial\zeta^2}\qquad\mathrm{with}\qquad
{\cal D}=\frac {3\pi^2} 2 \frac{R_\star\omega_\star^4}{v^2\omega^2}=\frac{3\pi^2}{2\ell(\omega)}\,;
\end{equation}
its boundary condition is $P(x=d,\zeta)=\delta(\zeta)$. The corresponding solution of Eq.~(\ref{eq:const-diff}) is
\begin{equation}
\label{eq:gaussian-distrib}
P(x=0,\zeta)=\frac1{2\sqrt{\pi {\cal D} d}}e^{-\zeta^2/(4{\cal D}d)}\,.
\end{equation}
The derivation of Eq.~(\ref{eq:gaussian-distrib}) required $\omega\gg\omega_\star$. Once this condition is satisfied, the obtained distribution function is valid at any ratio $d/\ell(\omega)$. [In particular, forward and backward scattering processes contribute additively (in a 2/3-1/3 ratio) to the ``diffusion constant'' $\cal D$.] This is quite remarkable, as the waves' localization length and mean free path are of the same order in a one-dimensional system. 

Averaging Eq.~\eqref{eq:res-rvdG3} over the distribution \eqref{eq:gaussian-distrib} then yields its disorder-averaged value at any $d$, see Eq.~(14) in the main text. Subsequently, one gets the disorder-averaged real part of the reflection amplitude through Eq.~(12) in the main text.

Using Kramers-Kronig relations, we get similarly the disorder-averaged real part of the local Green function,
\begin{equation}
\label{eq:r-im}
\frac{2\omega}{\pi^2 v}\text{Re}\langle G(0,0;\omega)\rangle = 2\nu_0
\sum_m\frac{ \sqrt{ 2}\Delta}{\pi \delta(\omega)} D\left(\frac{\omega-(m+1/2)\Delta}{\sqrt{2}\delta(\omega)}\right)\,,
\end{equation}
where $D(x)=e^{-x^2}\int_0^xdt e^{t^2}$ is the Dawson function, with asymptotes $D(x)\approx x$ at $|x|\ll 1$ and $D(x)\approx 1/(2x)$ at $|x|\gg 1$. In the frequency range $\omega_\star\ll\omega\ll \omega_\text{cr}$, the Poisson summation formula applied to Eq.~\eqref{eq:r-im} yields
\begin{equation}
\label{eq:r-im-2}
\frac{2\omega}{\pi^2 v} \text{Re}\langle G(0,0;\omega)\rangle \rangle = 4\nu_0\sin\left({2\pi \omega}/{\Delta}\right)e^{-2(\pi \omega_{\rm cr}/{\omega})^2}\,.
\end{equation}
Equations \eqref{eq:r-im} and \eqref{eq:r-im-2} determine the disorder-averaged imaginary part of the reflection amplitude,
\begin{equation}
\label{eq:local-dos}
\langle r''(\omega)\rangle =\frac{K}{K_0}\frac{2\omega}{\pi} \text{Re}\langle G(0,0;\omega)\rangle \,.
\end{equation}

\section{Soft modes}

Here we calculate the disordered-averaged local density of states at low frequency $\omega\ll\omega_\star$ in the strong-pinning regime. We confirm the scaling captured by Eq.~(17) in the main text, and we obtain the numerical factor in front of it.

Infinitely-strong pinning on separate impurities reduces the system to a sequence of independent segments. Averaging the local density of state in the segment nearby the interface at $x=0$, and whose length is given by the Poisson distribution, $P(L)=ce^{-cL}$, yields
\begin{equation}
\label{eq:DOS-Gorkov}
\langle \nu(x=0,\omega)\rangle
\approx 2c/\omega e^{-\pi v c/(2\omega)} \quad\text{at}\quad \omega\ll\pi v c\,.
\end{equation}

Finite-strength impurities allow for special configurations carrying soft excitations with arbitrarily-low frequency, as was noticed in~\cite{Aleiner1994}. Closely following their work, we consider three impurities at positions $x_1=L_1,x_2=x_1+L,x_3=x_2+L_2>0$ nearby $x=0$, and such that $L\ll 1/(\gamma c) \ll L_1,L_2$, where $\gamma=K\lambda/(\hbar v\sqrt{ac^3})$ is the large parameter in the strong pinning regime. The sum of elastic and potential energies associated with a static charge density such that $\partial_x\bar\theta(0)=0$ and $\bar\theta(x_j)=\theta_j$, 
\begin{equation}
E=\frac{\hbar v}{2\pi K}
\left[\frac{(\theta_2-\theta_1)^2}{L}+\frac{(\theta_3-\theta_2)^2}{L_2}\right]-\frac\lambda{\sqrt{ac}}\sum_{j=1}^3\cos(2\theta_j+\chi_j)\,,
\end{equation}
is minimized with $\theta_3=-\chi_3/2$ 
and $\theta_1-\theta_2=-2\pi  \gamma c L \sin[(\chi_1-\chi_2)/2]\cos \Phi \ll 1$ with $\Phi=\theta_1+\theta_2+(\chi_1+\chi_2)/2$. Then, if $\chi_1-\chi_2=\pi+2\varepsilon$ with $0<\varepsilon\ll 1$, the equilibrium solution for $\Phi$ (with $|\Phi|\ll 1$) is obtained from the minimization of a quartic potential,
\begin{equation}
\label{eq:E-quartic}
U=-h\Phi+A\Phi^2+B\Phi^4\,,
\end{equation}
where $h=\hbar v[\theta_3+(\chi_1+\chi_2)/4]/(2\pi K L_2)$, $A=\hbar v/(8\pi K L_2)-\varepsilon \lambda/\sqrt{ac}+ 4\pi K\lambda^2 L/(\hbar vac)$, and $B=\varepsilon \lambda/(12\sqrt{ac})-4\pi K\lambda^2 L/(3\hbar vac)$.

Using a Born-Oppenheimer approximation, we find the oscillation frequency $\omega_0$ in the minimum of the potential~\eqref{eq:E-quartic} using the Lagrangian $L=T-U$ where $T=M\dot\Phi^2/2$ with an effective ``mass'' $M=\hbar (L_1+L_2/3)/(\pi  vK)$ for the coordinate $\Phi$. Namely,
\begin{equation}
\label{eq:freq2}
\omega_0^2=\frac{4}M (h^2B)^{1/3} F\left( A/{(h^2B)^{1/3}}\right),
\end{equation}
where $F(\eta)$ is defined implicitly by $F(\eta)=\eta+6\xi^2$ with $-1+2\eta\xi+4\xi^3=0$. Function $F(\eta)$ reaches a minimum of order $1$ at $\eta=3/2^{5/3}$.

Using Eq.~\eqref{eq:freq2} and the properties of $F(\eta)$, we find that the phase space $\Gamma$ that allows for small oscillation frequencies $\omega_0\ll \omega$ is bounded by critical values for $|h|$ and $|A|$ scaling like $\omega^3$ and $\omega^2$, respectively. Thus, $\Gamma\propto \omega^5$, yielding a density of state $\nu\sim d\Gamma/d\omega\propto \omega^4$. Furthermore, the conditions $B>0$ and $A\approx 0$, which are necessary to find soft modes, yield two constraints: $\varepsilon<1/(6\pi L_2c\gamma)$ and $L<\varepsilon/(16\pi c\gamma)$, where $L_2\sim 1/c$. They yield an additional smallness in the suppression of the local density of states,
\begin{equation}
\label{eq:dos-universal}
\langle \nu(x=0,\omega)\rangle \propto \nu_0 \gamma^{-3} (\omega/vc)^4
\,.
\end{equation}
 The power-law \eqref{eq:dos-universal} dominates at low frequencies, while the exponential suppression \eqref{eq:DOS-Gorkov} takes over in an intermediate range of frequencies, $vc/\ln\gamma\ll \omega\ll vc$. 
 
The missing numerical factor in Eq.~(\ref{eq:dos-universal}) is found by an explicit calculation of the average $\langle\nu(x=0,\omega)\rangle$ at $\omega\ll\omega_\star$,

\begin{eqnarray}
\label{eq:dos-interm}
\langle \nu(x=0,\omega)\rangle=\int_0^\infty dL_1P(L_1) \int_0^\infty dL_2P(L_2) \int_0^{L_0} dL P(L)
\int_0^{2\pi} \frac{d\chi_1}{2\pi}\int_0^{2\pi} \frac{d\chi_2}{2\pi}\int_0^{2\pi} \frac{d\chi_3}{2\pi}\nonumber\\
\times \int_{-\infty}^\infty dh' \int_{-\infty}^\infty  dA'\delta(h-h')\delta(A-A')
\frac 1{L_1+L_2/3}\delta(\omega-\omega_0(A',h'))\,,
\end{eqnarray}
where the factor $1/(L_1+L_2/3)$ comes from the normalisation of the wavefunction associated with small oscillations. Here we used the Poisson distribution for the segments' lengths, and we assumed a uniform distribution for phases $\chi_1,\chi_2,\chi_3$ in the impurity potential. Furthermore, the upper limit $L_0=1/(96 \pi^2 L_2 c^2\gamma^2)$ in the integral over $L$ is set by the validity of expansion (\ref{eq:E-quartic}) leading to Eq.~(\ref{eq:freq2}). Now, setting $h'\approx A'\approx 0$ in the Dirac distributions (which is required to obtain low-frequency modes) allows performing the integrations over $h'$ and $A'$,
\begin{equation}
\int\!\!
dh' \int\!\!  
dA'
\delta(\omega-\omega_0(A',h'))=2\omega \int \!\!d\eta \int \!\!dh (h^2B)^{1/3}\delta\left(\omega^2-\frac{4(h^2B)^{1/3}}{M}F(\eta)\right)
=\frac 3{16}\frac{M^{5/2}\omega^4}{B^{1/2}}\int\!\! 
\frac{d\eta}{F(\eta)^{5/2}}
=\frac 1{8}\frac{M^{5/2}\omega^4}{B^{1/2}}\,.
\end{equation}
The remaining integrations in Eq.~\eqref{eq:dos-interm} can be performed to yield
\begin{eqnarray}
\langle \nu(x=0,\omega)\rangle&=&\frac 1\pi \nu_0\frac{1}{\gamma^2}\left(\frac\omega{vc}\right)^4c^5
\int_0^\infty dL_1\int_0^\infty dL_2 \int_0^{L_0} dL \frac{(L_1+L_2/3)^{3/2}L_2}{\sqrt{L_0-L}} e^{-c(L_1+L_2+L)}
\nonumber
\\
\nonumber\\
&=&\frac 1{2\sqrt{6}\pi^2}\nu_0\frac{1}{\gamma^3}\left(\frac\omega{vc}\right)^4 \int_0^\infty dx_1 dx_2 \sqrt{x_2(x_1+x_2/3)^3}e^{-x_1-x_2} 
\nonumber\\
&\approx& 0.12\, \nu_0\frac{1}{\gamma^3}\left(\frac\omega{vc}\right)^4\,,
\end{eqnarray}
where we used that the ranges of integration giving the dominant contributions are $L\ll 1/c$ and $L_1,L_2\sim 1/c$. The scaling is the same as in~\cite{Aleiner1994} for the global density of states, but the numerical prefactor is different.

At $\gamma \sim 1$, a crossover to the weak pinning occurs. That leaves no room for the intermediate asymptote \eqref{eq:DOS-Gorkov} and broadens the region for the $\omega^4$-dependence to $\omega\lesssim\omega_\star$, leading to 
\begin{equation}
\label{eq:Dos-w4}
\langle \nu(x=0,\omega)\rangle \propto \nu_0 (\omega/\omega_\star)^4
\,.
\end{equation}
The missing here, but appearing in Eq.~(17) of the main text, proportionality coefficient is found within a numerical simulation outlined in the next Section.

\section{Numerics}
\label{sec:numerics}

Here we provide details on the numerical computation of the local density of state at arbitrary frequency, which allowed us plotting Fig.~2 in the main text.

For this, we start by formulating the problem in dimensionless variables. Namely, by rescaling the spatial dimension $x=\xi y$ with length $\xi=R_\star/(2\pi^2)^{1/3}$,
we find that the static field $\bar\theta(y)$ setting the charge density in a given charge disorder configuration in the disconnected Josephson junction chain should admit boundary conditions $\partial_y\bar\theta(0)=0$ and $\bar\theta(d/\xi)=0$, and minimize the energy
\begin{equation} 
{\cal E}[\bar\theta]=\frac{\hbar v}{2\pi K\xi}\int_0^{d/\xi} dy\left[(\partial_y\bar\theta)^2-{\cal V}'\cos2\bar\theta+{\cal V}''\sin2\bar\theta\right]\,.
\end{equation}
Here ${\cal V}'(y)$ and ${\cal V}''(y)$ are two independent real random fields characterizing the disorder configuration, such that
\begin{equation}
\label{eq:correl-dimless}
\langle {\cal V}'(y){\cal V}'(y')\rangle=\langle {\cal V}''(y){\cal V}''(y')\rangle=\delta(y-y'')\,.
\end{equation}
The spectrum of small oscillations in the potential set by the static field $\bar\theta$ is also found as a dimensionless eigenproblem,
\begin{equation}
\label{eq:eigen-pb}
\Omega_n^2\psi_n=-\partial_y^2\psi_n+2({\cal V}'\cos2\bar\theta-{\cal V}''\sin2\bar\theta)\psi_n
\end{equation}
with eigenfrequency $\Omega_n=\omega_n \xi/v$ and (real) eigenfunction $\psi_n$ normalized such that ${\int_0^{d/\xi} dy\psi^2_n(y)}=1$.
The local density of states then reads as
\begin{equation}
\nu(x=0,\omega)=\pi\nu_0\sum_n {\psi^2_n(0)}\delta(\Omega-\Omega_n)\,.
\end{equation}

Next we discretize the above equations by introducing a small spacing $\epsilon$, such that the (dimensionless) length corresponds to $M=(d/\xi)/\epsilon$ sites. The random fields with correlators \eqref{eq:correl-dimless} are replaced with discrete fields ${\cal V}'_{m}$ and ${\cal V}''_{m}$ at sites $y_m=m\epsilon$ ($0\leq m<M$), which are uncorrelated from site to site, and which are drawn with flat probability in the interval $-\sqrt{3/\epsilon}<{\cal V}'_{m},{\cal V}''_{m}<\sqrt{3/\epsilon}$.

The random static field $\bar \theta$ is obtained by minimization of the energy functional
\begin{equation}
{\cal E}[\{\bar\theta_m\}]=\frac{\hbar v\epsilon}{2\pi K\xi}
\left\{
\sum_{m=1}^{M-1}\left[
\frac{(\bar\theta_{m-1}-\bar\theta_{m})^2}{\epsilon^2}
-({\cal V}'_{m}\cos2\bar\theta_m-{\cal V}''_{m}\sin2\bar\theta_m)\right]
+\frac{\bar\theta^2_{M-1}}{\epsilon^2}-\frac 12
({\cal V}'_{0}\cos2\bar\theta_0-{\cal V}''_{0}\sin2\bar\theta_0)
\right\}\,.
\end{equation}
The two last terms are such that the boundary conditions for $\bar \theta$ are satisfied in the continuum limit, $\epsilon\to 0$. (In particular, in order to reproduce the boundary condition at $y=0$ and get this result, one may think of $\bar\theta$ as the even solution of a twice longer chain with symmetric disorder on either side of the central node $m=0$.) We then use the algorithm described in Ref.~\cite{Gurarie2003} to obtain the variables $\{\bar\theta_m\}$ [taking discrete values $2\pi p/P$ ($0\leq p\leq P-1$ with integer $P\gg 1$)] that realize the absolute minimum of the energy $\cal E$.

Once the static configuration field has been determined, we can obtain the local density of states,
\begin{equation}
{\nu(x=0,\omega)}=-2{\nu_0}\Omega\text{Im}G_{0,0}(\Omega)\,,
\end{equation}
in terms of the retarded Green function $G_{m,m'}(\Omega)$ associated with Eq.~\eqref{eq:eigen-pb}. The boundary condition at site $m=0$ is obtained by considering a Green function $\tilde G_{m,m'}(\Omega)$ in a twice longer space,
\begin{equation}
G_{m,m'}=\tilde G_{m,m'}+\tilde G_{-m,m'}\,,
\end{equation}
such that $G_{0,0}=2\tilde G_{0,0}$. The later Green function is obtained as a continued fraction using the recursive equation 
\begin{equation}
(\Omega^2-{\cal W}_m+\frac 2{\epsilon^2})\tilde G_{m,m'}(\Omega)
-\frac 1{\epsilon^2}\left[\tilde G_{m+1,m'}(\Omega)+\tilde G_{m-1,m'}(\Omega)\right]=\delta_{m,m'}
\end{equation}
with ${\cal W}_m=2({\cal V}'_m\cos2\bar\theta_m-{\cal V}''_m\sin2\bar\theta_m)$.

In the plots shown here and in Fig.~2 in the main text, we choose a number of sites $M=200,400,\,\text{and}\,800$, and a mesh $\epsilon=0.05$, corresponding to $d/R_\star=M\epsilon /(2 \pi^{2})^{1/3}\approx 3.7,7.4,\,\text{and}\,14.8$, $\omega_\star/\Delta=n\epsilon /(2 \pi^{5})^{1/3}\approx 1.2,2.4,\,\text{and}\,4.7$, and $\omega_{\text{cr}} =\sqrt{3/2}(n\epsilon)^{3/2} /\pi^2\approx 3.9,11.1,\,\text{and}\,31.4$, respectively. We also introduce an imaginary broadening $\Omega\to\Omega+i\tilde \gamma$ with $\tilde \gamma=0.02 \Delta$.

\begin{figure}
(a)\includegraphics[width=0.35\columnwidth]{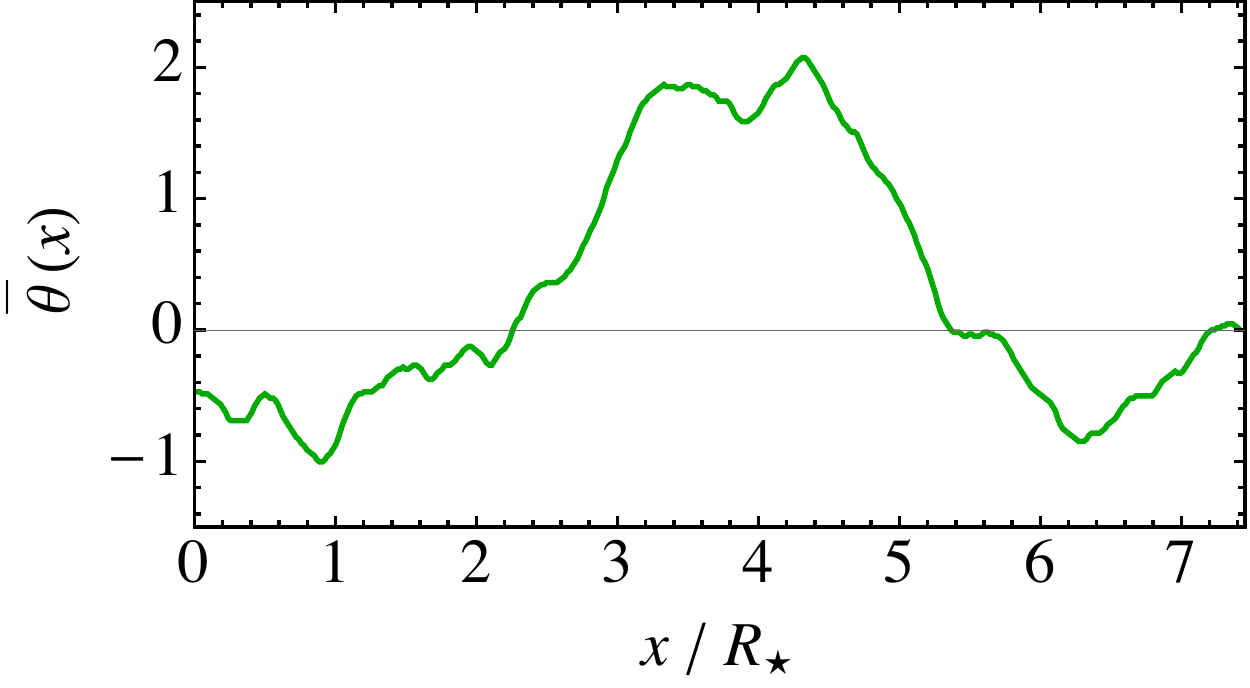}
(b)\includegraphics[width=0.34\columnwidth]{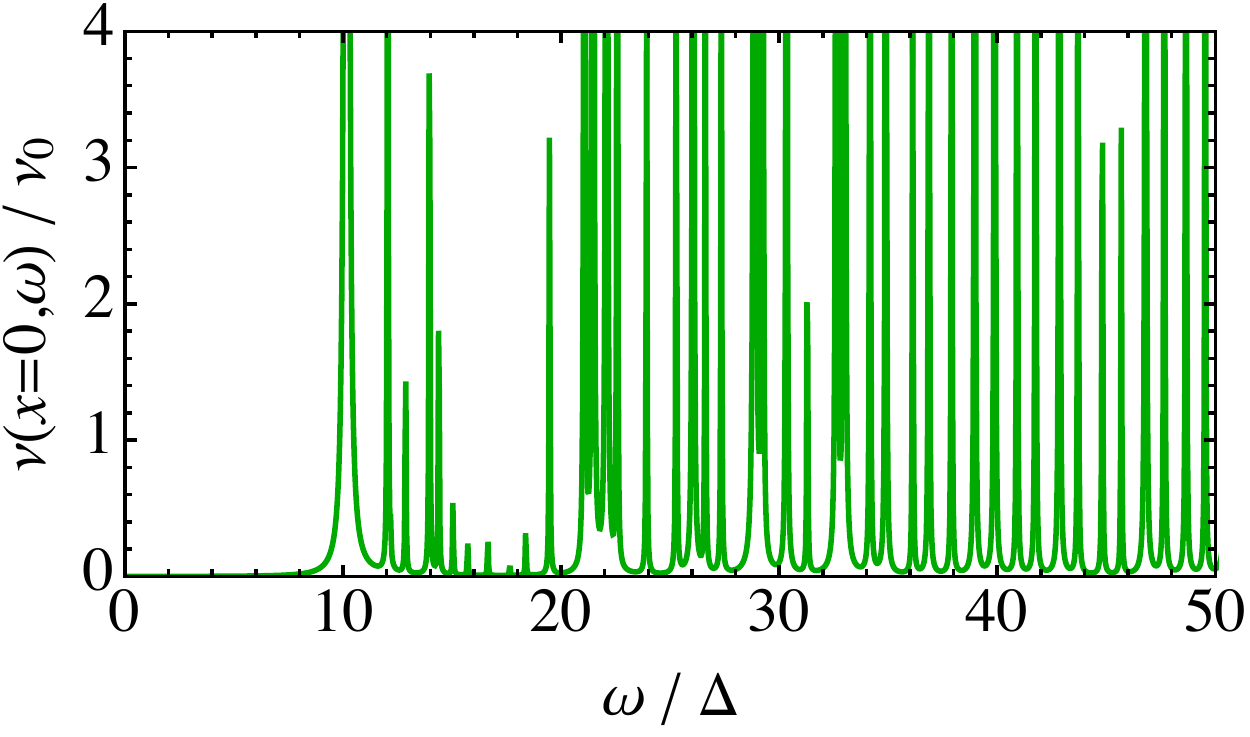}
\caption{\label{Fig:single-dis} (a) Spatial profile of the static configuration $\bar\theta(x)$ for a given disorder configuration in a chain with length $d=7.4 R_\star$, corresponding to $\omega_\star\approx 2.4 \Delta$ and $\omega_{\text{cr}} \approx 11.1 \Delta$. (b) Local density of states (in units of $\nu_0$) as a function of the frequency (in units of $\Delta$) for that disorder configuration.}
\end{figure} 

In Fig.~\ref{Fig:single-dis}, we plot the spatial profile of the static configuration along the chain for a single disorder configuration (left panel). We also show the local density of states (normalized by the bulk value $\nu_0$) as a function of the frequency (in units of $\Delta$) for that disorder configuration. Typically, the distant levels do contribute to the local density of states at low energies, but the contributions are exponentially small, $\exp(-d/R_\star)$. Thus one sees a few ``representative'' levels contributing to the local density of states at $\omega\lesssim\omega_\star$. By contrast, a (roughly) regular spectrum of equidistant states is visible at large frequencies  $\omega\gtrsim\omega_\text{cr}$.

\begin{figure}
(a)\includegraphics[width=0.35\columnwidth]{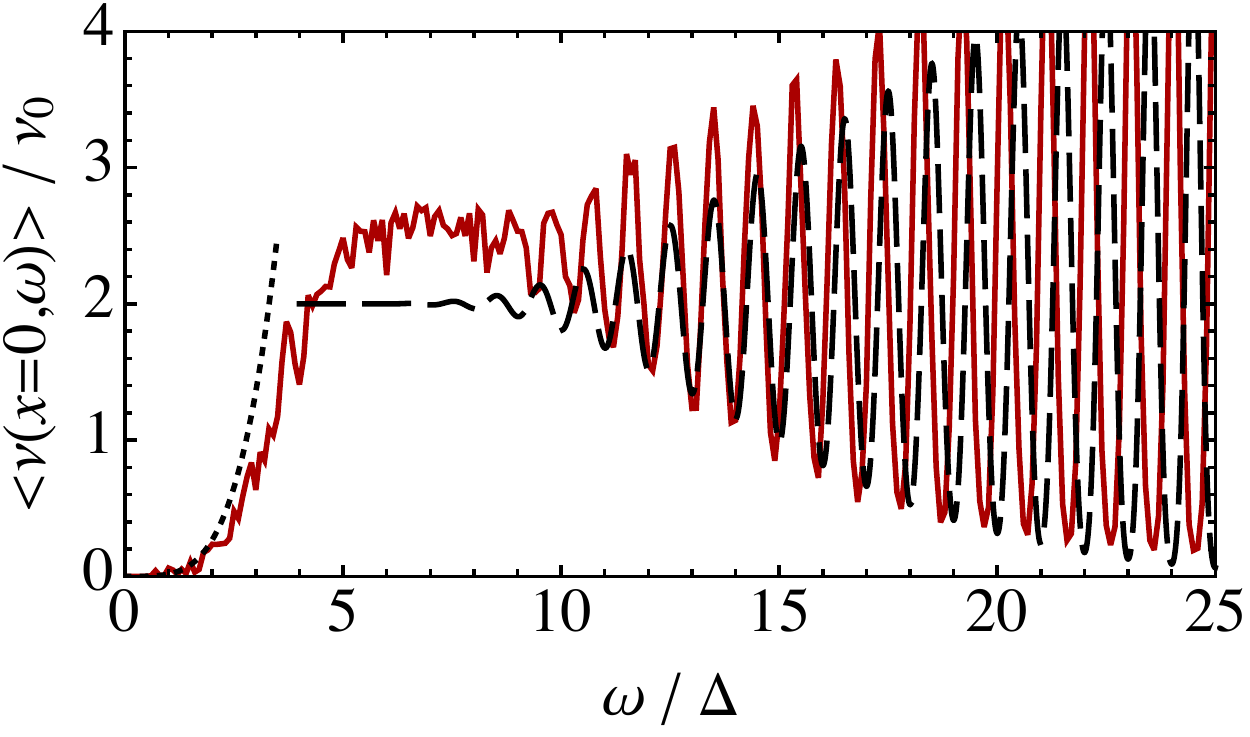}
(b)\includegraphics[width=0.35\columnwidth]{Fig-20.pdf}
\\
(c)\includegraphics[width=0.35\columnwidth]{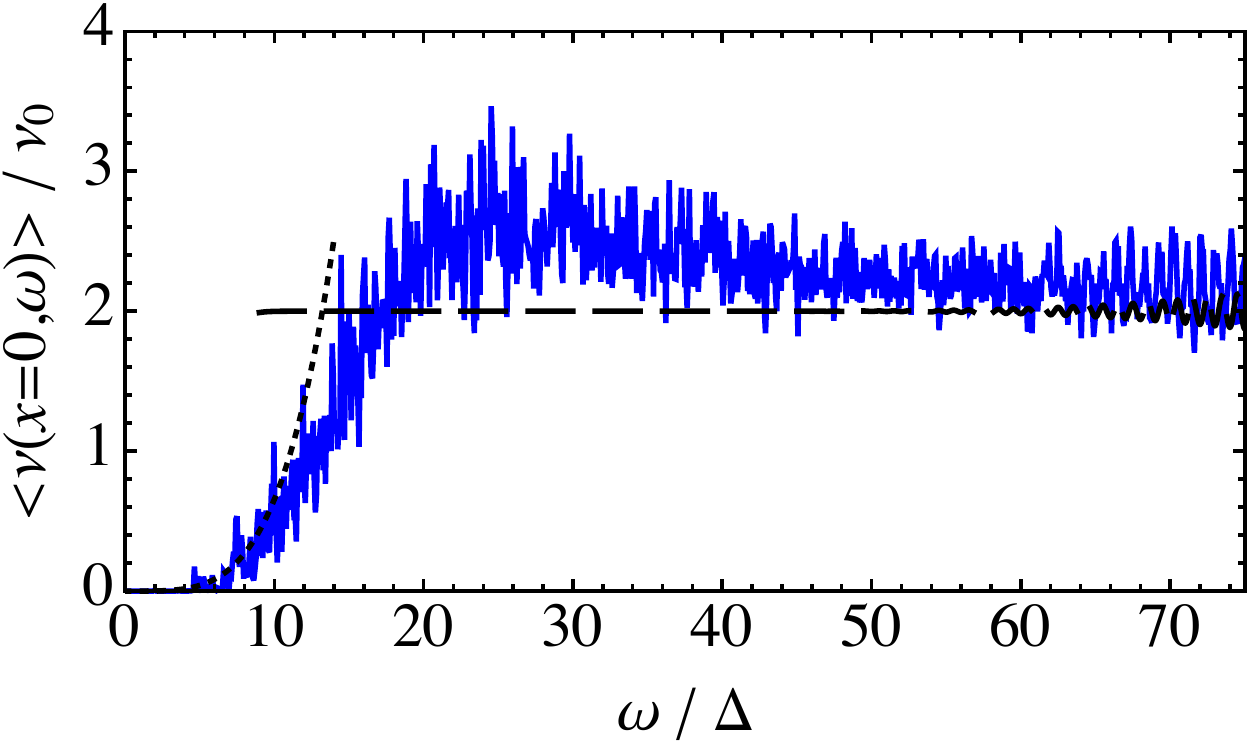}
(d)\includegraphics[width=0.35\columnwidth]{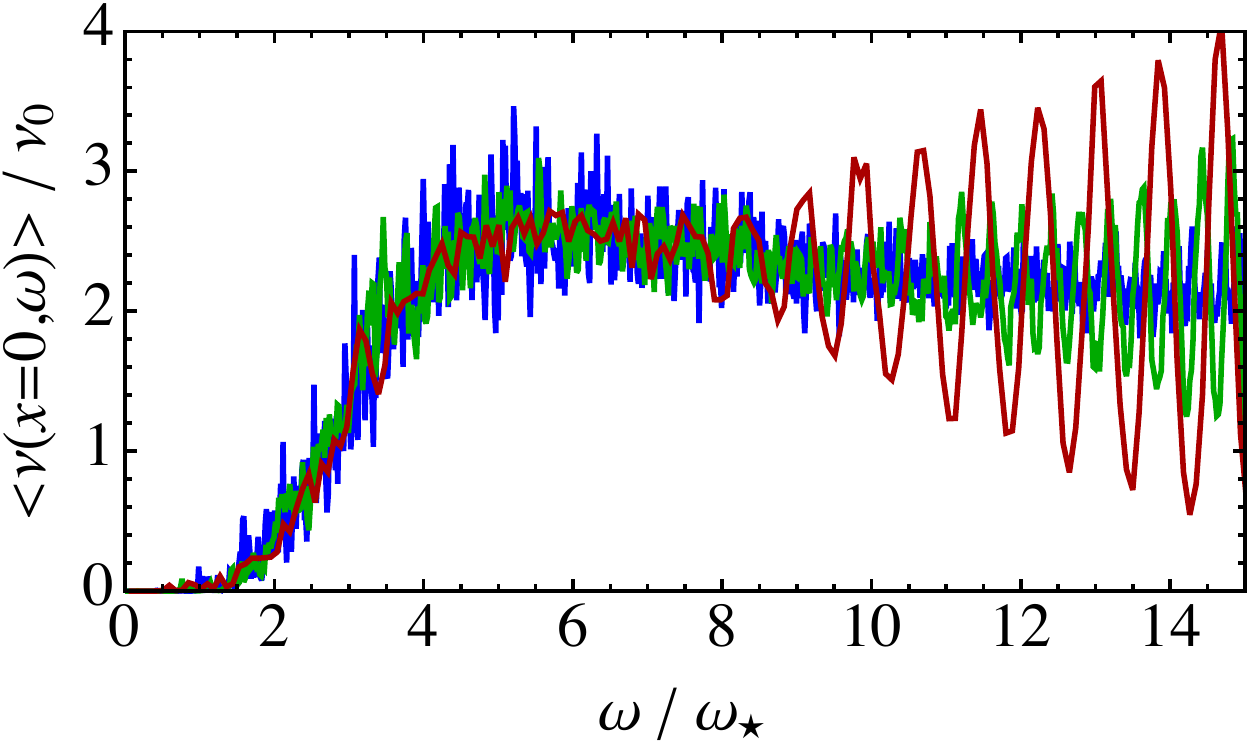}
\caption{\label{Fig:av-dis} (a-c) Local density of states (in units of $\nu_0$) as a function of the frequency (in units of $\Delta$) averaged over 5000 disorder configurations, and comparison with the asymptotic predictions at frequencies below and above $\omega_\star$, for chains with length $d=3.7,7.4,\,\text{and}\,14.8 R_\star$, respectively (Fig.~\ref{Fig:av-dis}(b) is same as Fig.~2 in the main text). (d) Disorder-averaged local density of states as a function of the frequency (in units of $\omega_\star$) for the three chain's lengths.} 
\end{figure} 

In Figs.~\ref{Fig:av-dis} and 2 in the main text, we plot the local density of states averaged over a large number of disorder configurations for three different lengths. We compare the result with the asymptotic formula
\begin{equation}
\langle \nu(0,\omega)\rangle /\nu_0= C(\omega/\omega_\star)^4 \quad \text{with}\quad C\sim 0.032\,,
\end{equation}
which was speculated to hold at low frequency $\omega\lesssim\omega_\star$, and with the prediction 
\begin{equation}
\langle \nu(0,\omega)\rangle/\nu_0 =
\sqrt{\frac 2 \pi}\frac{\omega_{\text{cr}}}{\omega}\sum_{n=0}^\infty \exp\left[-\frac{\omega^2}{2\omega_{\text{cr}}^2}\left(\frac \omega \Delta-n-\frac 12\right)^2\right]\,,
%[ 2 x/omega1/Sqrt[2 Pi] Exp[-x^2 (x - n - 1/2)^2/2/omega1^2]
\end{equation}
which holds at frequencies $\omega\gg\omega_\star$, yielding well resolved peaks at $\omega\gg\omega_\text{cr}$.
\begin{figure}
\includegraphics[width=0.35\columnwidth]{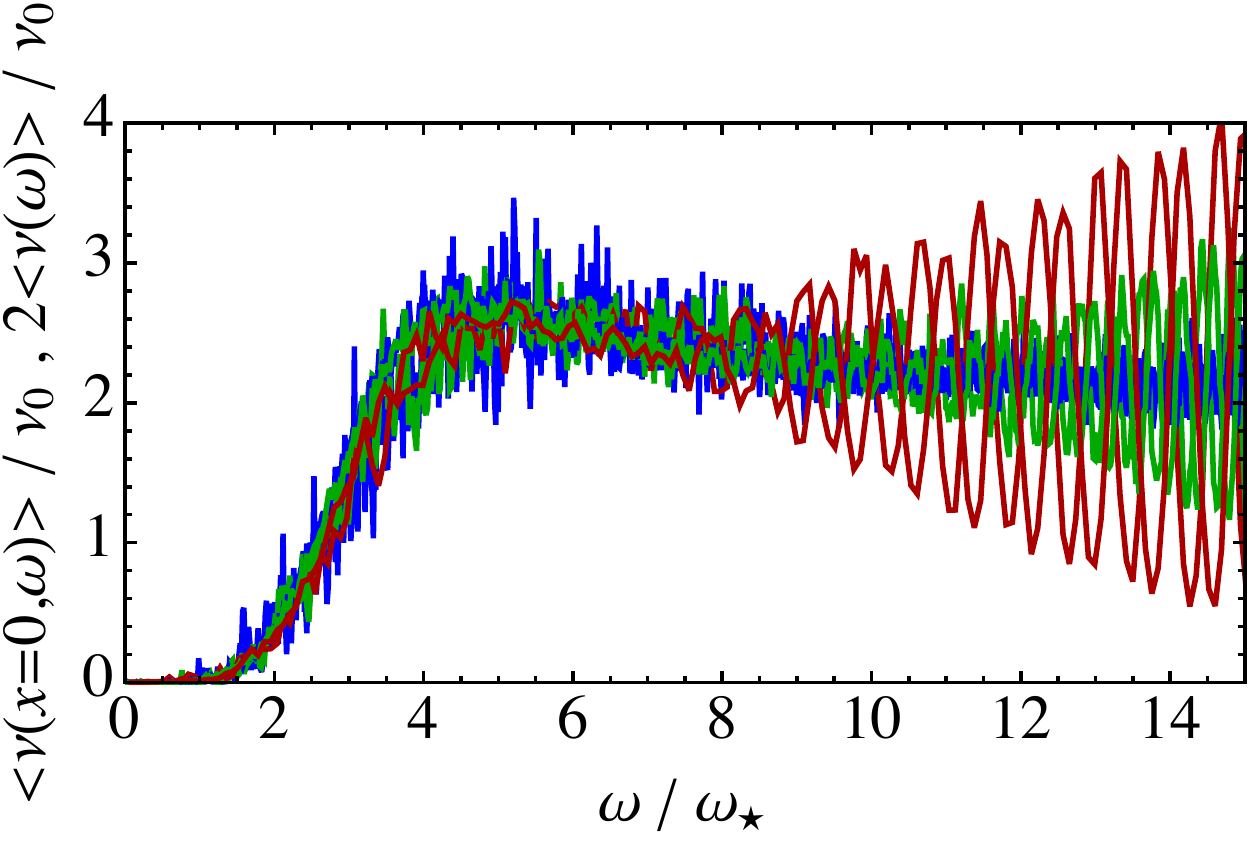}
\caption{\label{Fig:all} Local and global disorder-averaged density of states as a function of the frequency, for the same parameters as in Fig.~\ref{Fig:av-dis}.} 
\end{figure} 
In particular, Fig.~\ref{Fig:av-dis}(d) shows the scaling of the disorder-averaged local density of states as a function of $\omega/\omega_\star$ for three different lengths of the chain. 

In addition, Fig.~\ref{Fig:all} shows both the local and global densities of states; it illustrates the relation $\langle \nu(x=0,\omega)\rangle\approx 2 \langle \nu(\omega)\rangle$, which approximately holds at low frequencies, $\omega\lesssim\omega_\star$.

\section{Inelastic scattering}

Here we use perturbation theory in the phase slip term in Hamiltonian~(3) of the main text to provide an alternative derivation of the frequency shift, yielding Eq.~(16) in the main text in long chains, $d\gg R_\star$, at large frequencies, $\omega\gg\omega_\star$. Then we use the same approach to argue that inelastic scattering processes give negligible corrections to the result.\\

\subsection{Frequency shift}
Hamiltonian (2) in the main text, with boundary conditions $\partial_x\theta(x=0)=0$ and $\theta(x=d)=0$ in a finite-length chain, is diagonalized as
\begin{equation}
H_0= \sum_{n=0}^\infty (n+1/2)\Delta a^\dagger_na_n\,,
\end{equation}
where $a_n$ is an annihilation operator of a plasmon, such that
\begin{equation}
\label{eq:theta}
\theta(x)= \sum_{n=0}^\infty \sqrt{\frac{K}{n+1/2}}(a_n+ a_n^\dagger ) \cos (q_n x)
\end{equation}
with $q_n=\pi (n+1/2)/d$.

The dominant effect of a small but finite $\lambda$ is to shift the resonance frequencies, $\omega_n=(n+1/2)\Delta +\delta_n$. In first order in $\lambda$, the frequency shift of mode $n$ is the energy difference between no plasmon in the chain and the singly-occupied $n$-plasmon mode, $\delta_n=1/\hbar\left(\langle 1_n|H_1|1_n\rangle-\langle 0|H_1|0\rangle\right)$, where $|0\rangle$ is the vacuum state, $|1_n\rangle=a^\dagger_n|0\rangle$, and 
\begin{equation}
\label{eq:phase-slip-term}
H_1=-\frac\lambda a\int dx\cos(2\theta+\chi)
\end{equation}
is the phase slip term. Inserting Eq.~\eqref{eq:theta} in it, we get
\begin{eqnarray}
\delta_n
&=&-\frac\lambda {\hbar a} \mathrm{Re}\int dx\langle 0|\prod_m
\left( a_ne^{2i\sqrt{\frac{ K}{m+1/2}}( a_m+ a^\dagger_m)\cos (q_m x)} a_n^\dagger-
e^{2i\sqrt{\frac{ K}{m+1/2}}( a_m+ a^\dagger_m)\cos (q_m x)}\right)|0\rangle e^{i\chi(x)}
 \nonumber\\
&\approx&\frac{4K \lambda }{\hbar a(n+1/2)}\int _0^d dx\sin^2(q_n x)\exp\left(-{2 K}\sum_{m}\frac{\sin^2(q_m x)}{m+1/2}\right)\cos\chi(x)\,,
\end{eqnarray}
where we used $n\gg 1$ to expand the factor with $m=n$ up to lowest order in $K$, and to add back the (small) term with $m=n$ in the exponential. In long arrays, $d\gg R_\star$ (which can only happen if $K<3/2$, according to Eq.~(7) in the main text), the exponential factor can be evaluated by taking the continuum limit, $\sum_m\approx (d/\pi ) \int dq$, replacing $\sin^2(q_mx)\approx 1/2$, and using  $1/\ell_{\mathrm{sc}}$ and $1/R_\star (\gg 1/d)$ as the high and low momenta cutoffs of the log-divergent sum (this takes into account the fact that plasmons with frequency $\omega<\omega_\star$ are localized, and do not contribute to the sum). It yields
\begin{equation}
\label{eq:dn}
\delta_n
=\frac{4 K\lambda (R_\star/\ell_{\rm sc})^{-K} }{\hbar a(n+1/2)}\int _0^d dx\sin^2(q_n x)\cos\chi(x)\,.
\end{equation}
On average, $\langle\delta_n\rangle=0$. Using the correlator (4) in the main text, we find that its variance can be presented as
\begin{equation} 
\label{eq:var}
\frac{\langle\delta^2_{n}\rangle}{\Delta^2}
=\frac{3}{\pi^2 (n+1/2)^2}\left(\frac{d}{R_\star}\right)^3\,,
\end{equation}
in agreement with Eq.~(15) in the main text upon identification $\delta\left(\omega=(n+1/2)\Delta\right)\equiv{\langle\delta^2_{n}\rangle}^{1/2}$.

\subsection{Internal dissipation}
Considered in the harmonic approximation, the plasmon excitation spectrum of the Josephson-junction chain is broadened by radiative decay. At strong impedance mismatch with the waveguide, $K/K_0\ll 1$, the corresponding level width $\Gamma=(2v/d)K/K_0$  is small [for discussion of $\Gamma$, see the main text above Eq.~(10)]. Interaction between the plasmons may provide additional channels of ``internal dissipation''. In the absence of phase slips, charge disorder does not affect the dynamics of phases $\varphi_n$ in Eq.~(1) of the main text. The anharmonicity stemming from the expansion of $\cos(\varphi_n-\varphi_{n-1})$ beyond the second order in $\varphi_n-\varphi_{n-1}$, results in a momentum-conserving interaction between the plasmons. Because the plasmon spectrum $\omega(q)$ is convex, decay of a plasmon occurs only in collisions with other plasmons present in the system (a spontaneous decay is not allowed by the energy and momentum conservation laws), see, {\sl e.g.}, Ref.~\cite{Lin2013}. On the contrary, interaction stemming from the phase slip term \eqref{eq:phase-slip-term} allows a plasmon to decay into a number of plasmons of lower frequency and may provide an additional channel of ``internal dissipation''. Leaving aside a complex question regarding the possibility of many-body localization~\cite{Aleiner2006} of excitations of an isolated system at $\Gamma\to 0$, here we provide a crude estimate of the contribution of the internal dissipation to the plasmon line width. 

The decay rate of a state $|i\rangle$ of Hamiltonian $H_0$ due to a perturbation $H_1$ is given by the Fermi's Golden Rule,
\begin{equation}
\Gamma_i=2\pi\sum_f\left|\langle f|H_1|i\rangle\right|^2\delta(\varepsilon_i-\varepsilon_f)
={\rm Re}\int_{-\infty}^{\infty} dt e^{i0t} \langle i|e^{iH_0t}H_1e^{-iH_0t}H_1|i\rangle\,.
\end{equation}
Here $\varepsilon_i$ and $\varepsilon_f$ are, respectively, the energies of the initial state $|i\rangle$ and final states $|f\rangle$. We take a plasmon state $|i\rangle=a^\dagger _n|0\rangle$ with the (unperturbed) value of energy $\varepsilon_i= (n+1/2)\Delta$ as an approximate eigenstate of $H_0$. Inserting Eq.~\eqref{eq:theta} into the phase slip term \eqref{eq:phase-slip-term}, we then identify contributions leading to its decay into $p$ plasmons,
\begin{equation}
H_1=\sum_{p \geq 2}\sum_{n_1\dots n_p}V_{n,n_1\dots n_p}  a_{n_1}^\dagger\dots a_{n_p}^\dagger a_n
+\text{H.c.}+\text{other terms}
\end{equation}
with 
\begin{equation}
V_{n,n_1\dots n_p} 
=\frac{(2i)^{p+1}}{p!}\frac{\lambda(R_\star/\ell_{\rm sc})^{-K}}{a}\frac{K^{(p+1)/2}}{\sqrt{nn_1\dots n_pn}}\int_0^d dx\frac {e^{i\chi(x)}+(-1)^pe^{-i\chi(x)}}{2}\cos(q_{n_1} x)\dots \cos(q_{n_p} x) \cos(q_{n} x)\,.
\end{equation}
Here the factor $(R_\star/\ell_{\rm sc})^{-K}$ has the same origin as the one that appears in Eq.~\eqref{eq:dn}; furthermore we used $n_i+1/2\approx n_i$ at $n_i\gg 1$.
Then, we evaluate the inelastic decay rate induced by $H_1$ using Wick's theorem, 
\begin{eqnarray}
\Gamma_n&\approx&
\frac{\lambda^2}{a^2}
\left(\frac{R_\star}{\ell_{\rm sc}}\right)^{-2K}
\sum_p \frac{(4K)^{p+1}}{p!}
\sum_{n_1\dots n_p}
\frac 1{n_1\dots n_p n}
\left(\frac{da}{2^{p+2}}\right)
\int dt \langle 0|a_na^\dagger_n(t)a_{n_1}(t)\dots a_{n_{p}}(t)c_{n_{p}}^\dagger \dots a_{n_1}^\dagger a_{n}a_{n}^\dagger |0\rangle
\nonumber\\
&=&\frac{\lambda^2}{a^2}
\left(\frac{R_\star}{\ell_{\rm sc}}\right)^{-2K}
\sum_p \frac{(4K)^{p+1}}{p!}
\sum_{n_1\dots n_p}
\frac 1{n_1\dots n_p n}
\left(\frac{da}{2^{p+2}}\right)
\int dt iG_n^>(-t)iG_{n_1}^>(t)\dots iG_{n_p}^>(t)
\label{Gamman1}
\end{eqnarray}
with $iG_l^>(t)= \langle a_l(t)a^\dagger_l(0)\rangle$.
For estimates, we replaced here a factor which depends on the disorder configuration by a respective average:
\begin{equation}
\left(\frac{da}{2^{p+2}}\right)
=
\int_0^d dx \int_0^ddy \frac {e^{i\chi(x)}+(-1)^pe^{-i\chi(x)}}{2}\frac {e^{i\chi(y)}+(-1)^pe^{-i\chi(y)}}{2} \prod_{m=n_1,\dots,n_p,n} \cos q_m x \cos  q_m y \,.
\end{equation} 
Within the Rayleigh-Schr\"odinger perturbation theory, $G^>_{l}(t)\approx -i e^{-i\omega_l t}e^{-\Gamma_l |t|/2}$ with $\omega_l$ evaluated in the absence of phase slips, $\omega_l \to \varepsilon_l=(l+1/2)\Delta$, and the decay rate coming solely from radiation of the mode $i$  into the waveguide, $\Gamma_l\to \Gamma$. Assuming that the internal dissipation does occur, we modify $G^>_{l}(t)$ by allowing for $\Gamma_l\neq\Gamma$  and for $\omega_l$ corrected by the phase slips in the presence of a specific configuration of disorder. This way, Eq.~(\ref{Gamman1}) becomes a self-consistent equation for finding $\Gamma_n$:
\begin{equation}
\Gamma_n=
\frac {2 d K^2\lambda^2}a
\left(\frac{R_\star}{\ell_{\rm sc}}\right)^{-2K}
\sum_{p\geq 2} \frac{(2K)^{p-1}}{p!}
\sum_{n_1\dots n_p}
\frac 1{n_1\dots n_p n}
\frac{\Gamma_{n_1}+\dots+\Gamma_{n_p}+\Gamma_n}{(\omega_{n_1}+\dots+\omega_{n_p}-\omega_n)^2+\frac 14 (\Gamma_{n_1}+\dots+\Gamma_{n_p}+\Gamma_n)^2}
\,.
\label{Gamman2}
\end{equation}
The disorder-induced frequency shifts $\omega_l-\varepsilon_l$ are important, as they destroy multiple resonances which would occur for the equidistant unperturbed spectrum $\varepsilon_l$ appearing in the denominators of the summand of Eq.~(\ref{Gamman2}). It is clear from Eq.~(15) of the main text that the typical shifts $|\omega_l-\varepsilon_l|\sim\omega_{\rm cr}/l$ depend on $l$, obstructing the resonances unless the sum of $\Gamma_l$ in the denominators of  Eq.~(\ref{Gamman2}) exceeds ${\rm min}\{\Delta,\omega_{\rm cr}/l\}$. 

We are interested in the estimate of $\Gamma_n$ for modes with relatively high frequencies, $\varepsilon_n\gtrsim\omega_{\rm cr}$, at which $\omega_{\rm cr}/n\lesssim\Delta$. Our analysis of Eq.~(\ref{Gamman2}) at $K\ll 1$ points to $\Gamma_n\ll\omega_{\rm cr}/n$ in that frequency range, while $\Gamma_l\sim\Delta$ at $\omega_l\sim\sqrt{ K}\omega_{\rm cr}$ (here we assume that albeit being small, $K$ is large enough to allow for $\sqrt{K}\omega_{\rm cr}\gtrsim\omega_\star$, {\it i.e.}, $R_\star/d\lesssim K\ll 1$). At small $K$, the main contribution to the sum over $p$ comes from the first term ($p=2$) corresponding to a decay of the plasmon mode $n$ into another high-frequency mode $n_1\approx n$ accompanied by emission of one broadened plasmon ($\Gamma_l\gtrsim\Delta$) with frequency below $\sqrt{K}\omega_{\rm cr}$. That allows us to replace summation over $\{n\}$ in Eq.~(\ref{Gamman2}) by integration. Dispensing with a factor $\sim\ln(\sqrt{K}\omega_{\rm cr}/\omega_\star)$ and with a numerical factor which do not alter our conclusions, we obtain
\begin{equation}
\label{eq:inel}
\Gamma(\omega)\sim 
K\frac{\omega^3_\star}{\omega^2}
\sim \Delta \frac{K\omega^2_\text{cr}}{\omega^2}\quad\text{at}\quad \omega \gtrsim \sqrt{K}\omega_\text{cr}\,.
\end{equation}
Note that the frequency dependence here agrees with the one found in \cite{Rosenow2007}, as well as \cite{Bard2018} [cf.~the second line in their Eq.~(43)], at $K\ll 1$. Using Eq.~(\ref{eq:inel}) and Eq.~(15) of the main text, the ratio $\Gamma(\omega)/\delta(\omega)$ can be presented in the form
\begin{equation}
\frac{\Gamma(\omega)}{\delta(\omega)}\sim K\,\frac{\omega_{\rm cr}}{\omega}\,.
\label{Gamman3}
\end{equation}
This confirms that:\\ (i) at $\omega\gtrsim\omega_{\rm cr}$ the main contribution to the widths of the disorder-averaged plasmon resonances comes from the inhomogeneous broadening $\delta(\omega)$ considered in the main text, rather than from the internal dissipation;\\(ii) due to the internal dissipation, $\Gamma(\omega)$ may reach a value $\sim\Delta$ at low frequencies, $\omega\sim \sqrt{K}\omega_{\rm cr}$.